\documentclass[preprint,prd]{revtex4}
\usepackage{amssymb}
\usepackage{amsmath}
\usepackage{graphicx}

\input{tcilatex}

\begin{document}

\title[]{Contrasting Classical and Quantum Vacuum States in Non-Inertial
Frames}
\author{Timothy H. Boyer}
\affiliation{Department of Physics, City College of the City University of New York, New
York, New York 10031}

\begin{abstract}
Classical electron theory with classical electromagnetic zero-point
radiation (stochastic electrodynamics) is the classical theory which most
closely approximates quantum electrodynamics. Indeed, in inertial frames,
there is a general connection between classical field theories with
classical zero-point radiation and quantum field theories. \ However, this
connection does not extend to noninertial frames where the time parameter is
not a geodesic coordinate. \ Quantum field theory applies the canonical
quantization procedure (depending on the local time coordinate) to a
mirror-walled box, and, in general, each non-inertial coordinate frame has
its own vacuum state. \ In particular, there is a distinction between the
"Minkowski vacuum" for a box at rest in an inertial frame and a "Rindler
vacuum" for an accelerating box which has fixed spatial coordinates in an
(accelerating) Rindler frame. In complete contrast, the spectrum of random
classical zero-point radiation is based upon symmetry principles of
relativistic spacetime; in empty space, the correlation functions depend
upon only the geodesic separations (and their coordinate derivatives)
between the spacetime points. \ The behavior of classical zero-point
radiation in a noninertial frame is found by tensor transformations and
still depends only upon the geodesic separations, now expressed in the
non-inertial coordinates. \ It makes no difference whether a box of
classical zero-point radiation is gradually or suddenly set into uniform
acceleration; the radiation in the interior retains the same correlation
function except for small end-point (Casimir) corrections. \ Thus in
classical theory where zero-point radiation is defined in terms of geodesic 
separations, there is nothing physically comparable to the quantum distinction
between the Minkowski and Rindler vacuum states. \ It is also noted that
relativistic classical systems with internal potential energy must be
spatially extended and can not be point systems. Based upon the classical
analysis, it is suggested that the claimed heating effects of acceleration
through the vacuum may not exist in nature.
\end{abstract}

\maketitle

\section{Introduction}

Classical electron theory with classical electromagnetic zero-point
radiation (stochastic electrodynamics) is the classical theory which comes
closest to quantum electrodynamics.\cite{Reviews} \ However, there seems to
be little interest in the physical interpretations provided by this
classical theory. \ This lack of interest in the related classical theory
holds even when quantum theory ventures into untested areas involving
noninertial coordinate frames such as appear in connection with black holes
and acceleration through the vacuum. \ In this article, we illustrate the
contrasting classical and quantum interpretations surrounding vacuum
behavior in an inertial and in a noninertial (Rindler) frame. \ Although the
ideas are believed to have much wider implications, the illustrations here
focus on a massless relativistic scalar field in two spacetime dimensions in
flat spacetime.

There is a general connection between the classical and quantum field
theories in an inertial frame.\cite{Connection} However, this connection
does not extend to noninertial frames where the time parameter is not a
geodesic coordinate. \ Irrespective of the spacetime metric, quantum field
theory regards one box as good as another when applying the canonical
quantization procedure to a mirror-walled box. \ In general, each
non-inertial coordinate frame has its own vacuum state. \ In particular,
there is a distinction between the "Minkowski vacuum" for a box at rest in
an inertial frame and a "Rindler vacuum" for an accelerating box which has
fixed spatial coordinates in an (accelerating) Rindler frame. \ It has been
claimed\cite{Ref1} that the radiation in a box in the Minkowski vacuum which
is very gradually speeded up to become a box in uniform acceleration, will
end up in the Rindler vacuum state; on the other hand, if the box in the
Minkowski vacuum is suddenly accelerated, then the box will contain Rindler
quanta. \ This quantum situation is completely different from that found in
classical physics. In the first place, the spectrum of random classical
zero-point radiation is based upon symmetry principles of relativistic
spacetime; the spectrum is such as to give correlation functions which
depend only upon the geodesic separations (and their coordinate derivatives)
between the spacetime points. In an inertial frame, the zero-point radiation
spectrum is Lorentz invariant,\cite{rel} scale invariant, and conformal
invariant.\cite{conformal} \ The behavior of zero-point radiation in a
noninertial frame is found by tensor transformations to the non-inertial
coordinates. \ In particular, we can calculate the spectrum of classical
zero-point radiation in an accelerating box, and we find that, except for
small endpoint (Casimir) effects, the spectrum and correlation functions are
the same as observed by a Rindler observer accelerating through zero-point
radiation. \ It makes no difference whether or not the box of classical
zero-point radiation is gradually or suddenly set into uniform acceleration;
the radiation in the interior retains the same zero-point spectrum. \ In
classical theory where zero-point radiation is defined in terms of geodesic 
separations, there is nothing physically comparable to the quantum distinction
between the Minkowski and Rindler vacuum states.

The work presented here involves only the free scalar field in a box with
Dirichlet boundary conditions in one spatial dimension. \ Also, we will be
interested only in the large-box approximation and will not treat the
Casimir effects associated with a the discrete normal mode structure of the
box. \ We start out in an inertial frame. We review the determination of the
classical zero-point spectrum in the box and also the canonical quantization
procedure for the corresponding quantum scalar field in the same box. \ Then
we turn to the situation of thermal equilibrium in the box and note the
contrasting classical and quantum points of view for thermal radiation. \
All of this work confirms the general connection between classical and
quantum free fields in an inertial frame in two spacetime dimensions. \ This
connection was treated earlier in four spacetime dimensions for
electromagnetic fields\cite{Connection} and for scalar fields.\cite%
{Interpret} Next we turn to the situation for a coordinate frame undergoing
uniform proper acceleration through Minkowski spacetime (a Rindler frame). \
Quantum field theory introduces a canonical quantization in a box at rest in
a Rindler frame which parallels that in an inertial frame, without making
any adjustment because of the nongeodesic time coordinate involved in the
quantization. \ In complete contrast, classical theory takes the correlation
function for zero-point radiation as dependent only upon the geodesic
separations of the field points, with tensor coordinate transformations
between various coordinate frames. \ In the limit of a large Rindler-frame
box, the classical radiation inside the box is shown to agree exactly with
the empty-space zero-point radiation of an inertial frame. \ However, in the
limit of a large Rindler-frame box, the quantum vacuum remains distinct from
the quantum empty-space inertial vacuum. \ It is also emphasized that
relativistic classical systems with internal potential energy must be
spatially extended and can not be point systems. In contrast, systems used
within quantum theory are often described as small (point) systems.\cite%
{Ref2} \ Based upon the classical analysis, it is suggested that the claimed
"heating effects of acceleration through the vacuum" may not exist in nature.

\section{The Vacuum State in an Inertial Frame}

\subsection{Scalar Field in Two Spacetime Dimensions}

We will consider a relativistic massless scalar field $\phi $ which is a
function of $(ct,x)$ in an inertial frame with spacetime metric $%
ds^{2}=g_{\mu \nu }dx^{\mu }dx^{\nu }$, where the indices $\mu $ and $\nu $
run over 0 and 1, $x^{0}=ct$, $x^{1}=x$, and 
\begin{equation}
ds^{2}=c^{2}dt^{2}-dx^{2}  \label{e1}
\end{equation}%
The behavior of the field $\phi $ follows from the Lagrangian density $%
\mathcal{L}=(1/8\pi )\partial ^{\mu }\phi \partial _{\mu }\phi $
corresponding to\cite{Goldstein} 
\begin{equation}
\mathcal{L}=\frac{1}{8\pi }\left[ \frac{1}{c^{2}}\left( \frac{\partial \phi 
}{\partial t}\right) ^{2}-\left( \frac{\partial \phi }{\partial x}\right)
^{2}\right] .  \label{e2}
\end{equation}%
The wave equation $\partial _{\mu }[\partial \mathcal{L}/\partial (\partial
_{\mu }\phi )]=0$ for the field is 
\begin{equation}
\frac{1}{c^{2}}\left( \frac{\partial ^{2}\phi }{\partial t^{2}}\right)
-\left( \frac{\partial ^{2}\phi }{\partial x^{2}}\right) =0.  \label{e3}
\end{equation}%
The associated stress-energy-momentum tensor density $\mathcal{T}^{\mu \nu
}=[\partial \mathcal{L}/\partial (\partial _{\mu }\phi )]\partial ^{\nu
}\phi -g^{\mu \nu }\mathcal{L}$ gives the energy density $u$ as 
\begin{equation}
u=\mathcal{T}^{00}=\mathcal{T}^{11}=\frac{1}{8\pi }\left[ \frac{1}{c^{2}}%
\left( \frac{\partial \phi }{\partial t}\right) ^{2}+\left( \frac{\partial
\phi }{\partial x}\right) ^{2}\right] ,  \label{e4}
\end{equation}%
and the momentum density as 
\begin{equation}
\mathcal{T}^{01}=\mathcal{T}^{10}=-\frac{1}{4\pi c}\frac{\partial \phi }{%
\partial t}\frac{\partial \phi }{\partial x}.  \label{e5}
\end{equation}%
The energy $U$ in the field in a one-dimensional box extending from $x=a$ to 
$x=b$ is 
\begin{equation}
U=\!\int_{a}^{b}dx\frac{1}{8\pi }\left[ \frac{1}{c^{2}}\left( \frac{\partial
\phi }{\partial t}\right) ^{2}+\left( \frac{\partial \phi }{\partial x}%
\right) ^{2}\right] .  \label{e6}
\end{equation}

\subsection{Radiation Spectrum in a Box}

Both classical and quantum field theories start with the normal mode
structure of the radiation field in a box. We consider standing wave
solutions which vanish at the walls $x=a$ and $x=b$ of the box (Dirichlet
boundary conditions) so that a normalized normal mode can be written as 
\begin{equation}
\phi _{n}(ct,x)=f_{n}\left( \frac{2}{b-a}\right) ^{1/2}\sin \left[ \frac{%
n\pi }{b-a}(x-a)\right] \cos \left[ \frac{n\pi }{b-a}ct-\theta _{n}\right] .
\label{e18}
\end{equation}%
where $f_{n}$ is the amplitude of the normal mode. \ The radiation field in
the box can be written as a sum over all the normal modes 
\begin{equation}
\phi (ct,x)=\sum_{n=1}^{\infty }f_{n}\left( \frac{2}{b-a}\right) ^{1/2}\sin %
\left[ \frac{n\pi }{b-a}(x-a)\right] \cos \left[ \frac{n\pi }{b-a}ct-\theta
_{n}\right] ,  \label{e19}
\end{equation}%
where $\theta _{n}$ is an appropriate phase. From Eq.~(\ref{e4}) we find
that each mode $\phi _{n}(ct,x)$ has the time-average spatial energy density 
\begin{align}
u_{n}(x)& =\left\langle \frac{1}{8\pi }\left[ \frac{1}{c^{2}}\left( \frac{%
\partial \phi _{n}}{\partial t}\right) ^{2}+\left( \frac{\partial \phi _{n}}{%
\partial x}\right) ^{2}\right] \right\rangle _{\mathrm{time}}  \notag \\
& =\frac{1}{8\pi }\left( \frac{n\pi }{b-a}\right) ^{2}f_{n}^{2}\frac{2}{b-a}%
\{\sin ^{2}\left[ \frac{n\pi }{b-a}(x-a)\right] \left\langle \sin ^{2}\left[ 
\frac{n\pi }{b-a}ct-\theta _{n}\right] \right\rangle _{\mathrm{time}}  \notag
\\
& \quad +\cos ^{2}\left[ \frac{n\pi }{b-a}(x-a)\right] \left\langle \cos ^{2}%
\left[ \frac{n\pi }{b-a}ct-\theta _{n}\right] \right\rangle _{\mathrm{time}%
}\}  \notag \\
& =\frac{1}{8\pi }\left( \frac{n\pi }{b-a}\right) ^{2}\frac{f_{n}^{2}}{b-a},
\label{e20}
\end{align}%
which is uniform in space. The total mode energy $U_{n}$ found by
integrating over the length of the box is given by 
\begin{equation}
U_{n}=\frac{1}{8\pi }\left( \frac{n\pi }{b-a}\right) ^{2}f_{n}^{2}.
\label{e21}
\end{equation}%
where the wave amplitude $f_{n}$ must be determined by some additional
physical considerations.

\subsection{Canonical Quantization of the Quantum Scalar Field}

Classical and quantum theories take different points of view regarding the
vacuum radiation field. \ Quantum field theory follows the canonical
quantization procedure which rewrites the cosine time dependence in terms of
complex exponentials (the positive and negative frequency aspects) and
introduces annnihilation and creation operators $\overline{a}_{n},\overline{a%
}_{n}^{+}$ for each normal mode $n$ so that the field becomes an operator
field $\overline{\phi }(ct,x)$ with a vacuum energy 
\begin{equation}
U_{n}=(1/2)\hbar \omega =(1/2)\hbar cn\pi /(b-a)  \label{energy0}
\end{equation}%
per normal mode. Thus from Eqs. (\ref{e19}), (\ref{e21}), and (\ref{energy0}%
) the quantum field is 
\begin{eqnarray}
\overline{\phi }(ct,x) &=&\sum_{n=1}^{\infty }\left( 8\pi \hbar c\frac{(b-a)%
}{n\pi }\right) ^{1/2}\left( \frac{2}{b-a}\right) ^{1/2}\sin \left[ \frac{%
n\pi }{b-a}(x-a)\right]  \notag \\
&&\times \frac{1}{2}\left\{ \overline{a}_{n}\exp \left[ i\frac{n\pi }{b-a}ct%
\right] +\overline{a}_{n}^{+}\exp \left[ -i\frac{n\pi }{b-a}ct\right]
\right\}  \label{f22}
\end{eqnarray}%
Here the operator $\overline{a}_{n}$ annihilates the vacuum, $\overline{a}%
_{n}|0>=0,$ and the operator commutation relations are $[\overline{a}_{n},%
\overline{a}_{n}]=$ $[\overline{a}_{n}^{+},\overline{a}_{n}^{+}]=0,~[%
\overline{a}_{n},\overline{a}_{n}^{+}]=1.$

In the quantum vacuum state $|0>$ in the inertial frame, the two-point
vacuum expectation value which is symmetrized in operator order is easily
calculated and takes the form%
\begin{eqnarray}
&<&0|\frac{1}{2}\{\overline{\phi }(ct,x)\overline{\phi }(ct^{\prime
},x^{\prime })+\overline{\phi }(ct^{\prime },x^{\prime })\overline{\phi }%
(ct,x)\}|0>  \notag \\
&=&\sum_{n=1}^{\infty }\ \frac{4\hbar c}{n}\sin \!\left[ \frac{n\pi }{b-a}%
(x-a)\right] \sin \!\left[ \frac{n\pi }{b-a}(x^{\prime }-a)\right] \cos \!%
\left[ \frac{n\pi }{b-a}c(t-t^{\prime })\right]  \label{f23}
\end{eqnarray}

\subsection{Zero-Point Radiation for the Classical Scalar Field}

The vacuum state for the classical scalar field involves random classical
zero-point radiation which is featureless, so that its correlation functions
depend only on the geodesic separations (and coordinate derivatives) between
the field points. \ In an inertial frame, the zero-point radiation is
Lorentz invariant,\cite{rel} scale invariant, and indeed conformal invariant.%
\cite{conformal} \ Random classical radiation can be written in the form
given by Eq. (\ref{e19}) with the phases $\theta _{n}$ randomly distributed
in the interval $[0,2\pi )$ and independently distributed for each $n$. \ In
an inertial frame, the invariance properties of the spectrum can be shown to
lead to a spectral form corresponding to an energy \ per normal mode which
is a multiple of the frequency with an undetermined multiplicative constant, 
$U_{n}=const\times \omega _{n}$.\cite{conformal} \ In order to give a close
connection between the classical and quantum theories, we choose the energy
per normal mode to agree with that used in the quantum theory as given in
Eq. (\ref{energy0}). \ In order to make the classical and quantum field
expressions look as similar as possible, we rewrite Eq. (\ref{e19}) in the
form parallel to Eq. (\ref{f22}), (note the change from $8\pi $ over to $%
4\pi ),$%
\begin{eqnarray}
\phi _{0}(ct,x) &=&\sum_{n=1}^{\infty }\left( 4\pi \hbar c\frac{(b-a)}{n\pi }%
\right) ^{1/2}\left( \frac{2}{b-a}\right) ^{1/2}\sin \left[ \frac{n\pi }{b-a}%
(x-a)\right] \cos \left[ \frac{n\pi }{b-a}ct-\theta _{n}\right]  \notag \\
&=&\sum_{n=1}^{\infty }\left( 4\pi \hbar c\frac{(b-a)}{n\pi }\right)
^{1/2}\left( \frac{2}{b-a}\right) ^{1/2}\sin \left[ \frac{n\pi }{b-a}(x-a)%
\right]  \notag \\
&&\times \frac{1}{2}\left\{ e^{-i\theta _{n}}\exp \left[ i\frac{n\pi }{b-a}ct%
\right] +e^{i\theta _{n}}\exp \left[ -i\frac{n\pi }{b-a}ct\right] \right\}
\label{f24}
\end{eqnarray}

It is convenient to characterize random classical radiation by the two-point
correlation function $\left\langle \phi (ct,x)\phi (ct^{\prime },x^{\prime
})\right\rangle $ obtained by averaging over the random phases as $%
\left\langle \cos \theta _{n}\sin \theta _{n^{\prime }}\right\rangle =0,$ $%
\left\langle \cos \theta _{n}\cos \theta _{n^{\prime }}\right\rangle
=\left\langle \sin \theta _{n}\sin \theta _{n^{\prime }}\right\rangle
=(1/2)\delta _{n,n^{\prime }},$ or as $\left\langle \exp [\theta _{n}]\exp
[\theta _{n^{\prime }}]\right\rangle =\left\langle \exp [-\theta _{n}]\exp
[-\theta _{n^{\prime }}]\right\rangle =0,$ $\left\langle \exp [\theta
_{n}]\exp [-\theta _{n^{\prime }}]\right\rangle =\delta _{n,n^{\prime }}$
From these relations, we can easily show, for example, that $\langle \cos
(A+\theta _{n})\cos (B+\theta _{n^{\prime }})\rangle =\cos (A-B)(1/2)\delta
_{nn^{\prime }}$. The two-point correlation function for a general
distribution of random classical scalar waves is found by averaging over the
random phases $\theta _{n}$ 
\begin{align}
& \left\langle \phi _{0_{box}}(ct,x)\phi _{0_{box}}(ct^{\prime },x^{\prime
})\right\rangle  \notag \\
& =<\sum_{n=1}^{\infty }\left( 4\pi \hbar \frac{{}}{{}}c\frac{(b-a)}{n\pi }%
\right) ^{1/2}\left( \frac{2}{b-a}\right) ^{1/2}\sin \!\left[ \frac{n\pi }{%
b-a}(x-a)\right] \cos \!\left[ \frac{n\pi }{b-a}ct-\theta _{n}\right]  \notag
\\
& \quad \times \sum_{n^{\prime }=1}^{\infty }\left( 4\pi \hbar c\frac{(b-a)}{%
n\pi }\right) ^{1/2}\left( \frac{2}{b-a}\right) ^{1/2}\sin \!\left[ \frac{%
n^{\prime }\pi }{b-a}(x^{\prime }-a)\right] \cos \!\left[ \frac{n^{\prime
}\pi }{b-a}ct^{\prime }-\theta _{n^{\prime }}\right] >  \notag \\
& =\sum_{n=1}^{\infty }\ \frac{4\hbar c}{n}\sin \!\left[ \frac{n\pi }{b-a}%
(x-a)\right] \sin \!\left[ \frac{n\pi }{b-a}(x^{\prime }-a)\right] \cos \!%
\left[ \frac{n\pi }{b-a}c(t-t^{\prime })\right] .  \label{e24}
\end{align}%
We notice that the classical correlation function (\ref{e24}) and vacuum
expectation value of (symmetrized) quantum operators (\ref{f23}) agree
exactly. \ Indeed it has been shown that in an inertial frame, there is a
general connection\cite{Connection} between the correlation functions of the
classical zero-point radiation field and the vacuum expectation values of
the corresponding symmetrized operator products for all the correlation
functions including the correlation functions of arbitrarily high order.

If we take the limit $b\rightarrow \infty ,$ corresponding to the presence
of a reflecting mirror at the left-hand end $x=a$ of the box but infinite
extent on the right, then we obtain the correlation function as an integral
where the wave numbers $k_{n}=n\pi /(b-a)$ become continuous,%
\begin{equation}
\quad \left\langle \phi _{0_{mirror}}(ct,x)\phi _{0_{mirror}}(ct^{\prime
},x^{\prime })\right\rangle =4\hbar c\dint_{k=0}^{k=\infty }\frac{dk}{k}\sin %
\left[ k(x-a)\right] \sin [k(x^{\prime }-a)]\cos \left[ kc(t-t^{\prime })%
\right] .  \label{f26}
\end{equation}%
This integral is convergent. \ It can be rewritten as a sum of terms of the
form $\tint (dk/k)\cos (ka)$ and evaluated as an indefinite integral. \ Thus
we find%
\begin{eqnarray}
\left\langle \phi _{0_{mirror}}(ct,x)\phi _{0_{mirror}}(ct^{\prime
},x^{\prime })\right\rangle &=&-\hbar c\ln \left\vert \frac{[(x-x^{\prime
})-c(t-t^{\prime })][(x-x^{\prime })-c(t-t^{\prime })]}{[(x+x^{\prime
}-2a)-c(t-t^{\prime })][(x+x^{\prime }-2a)-c(t-t^{\prime })]}\right\vert 
\notag \\
&=&-\hbar c\ln \left\vert \frac{(x-x^{\prime })^{2}-c^{2}(t-t^{\prime })^{2}%
}{(x+x^{\prime }-2a)^{2}-c^{2}(t-t^{\prime })^{2}}\right\vert  \label{f26a}
\end{eqnarray}

The correlation function for empty space can be found by moving the mirror
at the left-hand edge $x=a$ of the box out to spatial infinity, $%
a\rightarrow -\infty .$ \ However, this procedure introduces a divergence
going as $\hbar c\ln \left\vert (2a)^{2}\right\vert .$ \ One way to
eliminate this divergence is to take the spatial derivatives of the
correlation function. \ Indeed, we can go back to the integral of Eq. (\ref%
{f26}) and use the identity $2\sin A\sin B=\cos (A-B)-\cos (A+B)$ to rewrite
the correlation function as 
\begin{eqnarray}
\left\langle \phi _{0_{mirror}}(ct,x)\phi _{0_{mirror}}(ct^{\prime
},x^{\prime })\right\rangle &=&2\hbar c\dint_{k=0}^{k=\infty }\frac{dk}{k}%
\cos \left[ k(x-x^{\prime })\right] \cos \left[ kc(t-t^{\prime })\right] 
\notag \\
&&-2\hbar c\dint_{k=0}^{k=\infty }\frac{dk}{k}\cos \left[ k(x+x^{\prime }-2a)%
\right] \cos \left[ kc(t-t^{\prime })\right]  \label{f26b}
\end{eqnarray}%
Both integrals in Eq. (\ref{f26b}) are divergent at $k\rightarrow 0$. \ In
the limit $a\rightarrow -\infty ,$ corresponding to moving the\ left-hand
reflecting mirror at $x=a$ out to spatial minus infinity, we can drop the
second line in Eq. (\ref{f27}) as a very rapidly oscillating cosine
function. \ Thus for a box extending infinitely far in both directions, we
find the free-space correlation function%
\begin{eqnarray}
\left\langle \phi _{0}(ct,x)\phi _{0}(ct^{\prime },x^{\prime })\right\rangle
&=&2\hbar c\dint_{k=0}^{k=\infty }\frac{dk}{k}\cos \left[ k(x-x^{\prime })%
\right] \cos \left[ kc(t-t^{\prime })\right]  \notag \\
&=&\hbar c\dint_{k=0}^{k=\infty }\frac{dk}{k}\cos \left[ k\{(x-x^{\prime
})+c(t-t^{\prime })\}\right]  \notag \\
&&+\hbar c\dint_{k=0}^{k=\infty }\frac{dk}{k}\cos \left[ k\{(x-x^{\prime
})-c(t-t^{\prime })\}\right]  \notag \\
&=&\hbar c\dint_{k=-\infty }^{k=\infty }\frac{dk}{|k|}\cos [k(x-x^{\prime
})-|k|c(t-t^{\prime })]  \label{f27}
\end{eqnarray}%
where in the second line and third lines we have used the identity $2\cos
A\cos B=\cos (A+B)+\cos (A-B)$ and in the last line have incorporated both
the sum and difference cosine terms by extending the integral over negative
values of $k$. \ 

The integrals in Eqs. (\ref{f26b}) and (\ref{f27}) are divergent as $%
k\rightarrow 0.$ \ This divergence can be removed by considering the
coordinate derivatives of the correlation functions. \ Thus in free space,
we consider $\left\langle \phi _{0}(ct,x)\partial _{ct^{\prime }}\phi
_{0}(ct^{\prime },x^{\prime })\right\rangle $ and $\left\langle \phi
_{0}(ct,x)\partial _{x^{\prime }}\phi _{0}(ct^{\prime },x^{\prime
})\right\rangle .$ \ The resulting expressions are convergent as $%
k\rightarrow 0$ but now divergent as $k\rightarrow \infty .$ \ However, the
divergence at large values of $k$ involves oscillating sine functions. \
Thus we may introduce a convergence factor such as $\exp [-\Lambda k]$ into
the integrand, carry out the integrals in terms of exponentials, and then
take the no-cutoff limit $\Lambda \rightarrow 0$ to obtain the singular
Fourier sine transforms of the form\cite{twoD}%
\begin{equation}
\dint_{0}^{\infty }dk\,k^{2m}\sin (bk)=\frac{(-1)^{2m}(2m)!}{b^{2m+1}}
\label{f27a}
\end{equation}%
In this fashion we obtain the closed-form expression 
\begin{eqnarray}
&&\frac{\partial }{\partial ct^{\prime }}\left\langle \phi _{0}(ct,x)\phi
_{0}(ct^{\prime },x^{\prime })\right\rangle  \notag \\
&=&\hbar c\dint_{k=0}^{k=\infty }dk\sin \left[ k\{(x-x^{\prime
})+c(t-t^{\prime })\}\right] -\hbar c\dint_{k=0}^{k=\infty }dk\sin \left[
k\{(x-x^{\prime })-c(t-t^{\prime })\}\right]  \notag \\
&=&\hbar c\left[ (x-x^{\prime })+c(t-t^{\prime })\right] ^{-1}-\hbar c\left[
(x-x^{\prime })-c(t-t^{\prime })\right] ^{-1}  \notag \\
&=&\frac{\partial }{\partial ct^{\prime }}\left\{ -\ln \left\vert
(x-x^{\prime })+c(t-t^{\prime })\right\vert -\ln \left\vert (x-x^{\prime
})-c(t-t^{\prime })\right\vert \right\}  \notag \\
&=&\frac{\partial }{\partial ct^{\prime }}\left\{ -\hbar c\ln \left\vert
c^{2}(t-t^{\prime })^{2}-(x-x^{\prime })^{2}\right\vert \right\} =2\hbar c%
\frac{c(t-t^{\prime })}{c^{2}(t-t^{\prime })^{2}-(x-x^{\prime })^{2}}
\label{f28}
\end{eqnarray}%
and similarly obtain%
\begin{eqnarray}
\frac{\partial }{\partial x^{\prime }}\left\langle \phi _{0}(ct,x)\phi
_{0}(ct^{\prime },x^{\prime })\right\rangle &=&\frac{\partial }{\partial
x^{\prime }}\left\{ -\hbar c\ln \left\vert c^{2}(t-t^{\prime
})^{2}-(x-x^{\prime })^{2}\right\vert \right\}  \notag \\
&=&2\hbar c\frac{-(x-x^{\prime })}{c^{2}(t-t^{\prime })^{2}-(x-x^{\prime
})^{2}}  \label{f29}
\end{eqnarray}%
both of which agree with the limit $a\rightarrow -\infty $ in Eq. (\ref{f26a}%
). \ We note that in empty space there is no length or time parameter which
is singled out by the zero-point radiation in an inertial frame. The
zero-point correlation functions depend upon the geodesic separation $%
c^{2}(t-t^{\prime })^{2}-(x-x^{\prime })^{2}$ between the field points $%
(ct,x)$ and $(ct^{\prime},x^{\prime })$.

For later comparisons, it is useful to have the closed-form expressions for
the zero-point correlation functions in empty space as a function of time at
a single spatial coordinate $x=x^{\prime }\,\ $and as a function of space at
a single time $t=t^{\prime }.$ \ Thus we have for the non-vanishing
correlations from Eqs. (\ref{f28}) and (\ref{f29})%
\begin{equation}
\left\langle \phi _{0}(ct,x)\partial _{ct^{\prime }}\phi _{0}(ct^{\prime
},x^{\prime })\right\rangle _{x^{\prime }=x}=2\hbar c\frac{1}{c(t-t^{\prime
})}  \label{f32}
\end{equation}%
and 
\begin{equation}
\left\langle \phi _{0}(ct,x)\partial _{x^{\prime }}\phi _{0}(ct^{\prime
},x^{\prime })\right\rangle _{t=t^{\prime }}=2\hbar c\frac{1}{(x-x^{\prime })%
}  \label{f33}
\end{equation}%
The spatial derivatives of the correlation function for a mirror at $x=a$ at
the left-hand end of the spatial region can be written explicitly as 
\begin{eqnarray}
&&\left\langle \phi _{0_{mirror}}(ct,x)\partial _{ct^{\prime }}\phi
_{0_{mirror}}(ct^{\prime },x^{\prime })\right\rangle  \notag \\
&=&2\hbar c\frac{\ c(t-t^{\prime })}{c^{2}(t-t^{\prime })^{2}-(x-x^{\prime
})^{2}}-2\hbar c\frac{\ c(t-t^{\prime })}{c^{2}(t-t^{\prime
})^{2}-(x+x^{\prime }-2a)^{2}}  \notag \\
&=&\frac{\partial }{\partial ct^{\prime }}\left\{ -\hbar c\ln \left\vert
c^{2}(t-t^{\prime })^{2}-(x-x^{\prime })^{2}\right\vert \right\} -\frac{%
\partial }{\partial ct^{\prime }}\left\{ -\hbar c\ln \left\vert
c^{2}(t-t^{\prime })^{2}-(x+x^{\prime }-2a)^{2}\right\vert \right\}
\label{f30}
\end{eqnarray}%
\begin{eqnarray}
&&\left\langle \phi _{0_{mirror}}(ct,x)\partial _{x^{\prime }}\phi
_{0_{mirror}}(ct^{\prime },x^{\prime })\right\rangle  \notag \\
&=&2\hbar c\frac{-(x-x^{\prime })}{c^{2}(t-t^{\prime })^{2}-(x-x^{\prime
})^{2}}+2\hbar c\frac{-(x+x^{\prime }-2a)}{c^{2}(t-t^{\prime
})^{2}-(x+x^{\prime }-2a)^{2}}  \notag \\
&=&\frac{\partial }{\partial x^{\prime }}\left\{ -\hbar c\ln \left\vert
c^{2}(t-t^{\prime })^{2}-(x-x^{\prime })^{2}\right\vert \right\} -\frac{%
\partial }{\partial x^{\prime }}\left\{ -\hbar c\ln \left\vert
c^{2}(t-t^{\prime })^{2}-(x+x^{\prime }-2a)^{2}\right\vert \right\}
\label{f31}
\end{eqnarray}

\subsection{Thermal Scalar Radiation}

Within classical theory with classical zero-point radiation, zero-point
radiation represents real radiation which is always present, and thermal
radiation is additional random radiation above the zero-point value. Thus if 
$U(\omega ,T)$ is the energy per normal mode at frequency $\omega $ and
temperature $T$, the thermal energy contribution $U_{T}(\omega ,T)$ is found
by subtracting off the zero-point energy, $U_{T}(\omega ,T)=U(\omega
,T)-U(\omega ,0).$ The additional thermal energy is distributed across the
low-frequency modes of the radiation field. \ The (finite) total thermal
energy $\mathcal{U}_{T}(T)$ in a box is found by summing the thermal energy
per normal mode $U_{T}(\omega ,T)$ over all the normal modes at temperature $%
T$ in a box of finite size. \ The spatial density of thermal energy is given
by $u(T)=\mathcal{U}_{T}(T)/(b-a)=a_{Ss}T^{2}$ where $a_{Ss}$ is the
constant for one-spatial-dimension scalar radiation corresponding to
Stefan's constant for electromagnetic radiation.\cite{twoD} \ Classical
thermal radiation is described in exactly the same random-phase fashion as
the zero-point radiation except that the spectrum is changed. \ The thermal
radiation spectrum for massless scalar radiation can be derived from
classical theory involving zero-point radiation and the structure of
spacetime.\cite{twoD}\cite{Derivation}\cite{structure} \ One \ finds for the
energy per normal mode at frequency $\omega $ and temperature $T$ 
\begin{equation}
U(\omega ,T)=(1/2)\hbar \omega \coth [\hbar \omega /(2k_{B}T)]  \label{f34}
\end{equation}%
The calculation for the classical two-point field correlation function at
finite temperature accordingly takes exactly the same form as given above in
Eqs. (\ref{e24}), except that the spectrum is changed so that now 
\begin{eqnarray}
&&\left\langle \phi _{T\,box}(ct,x)\phi _{T\,box}(ct^{\prime },x^{\prime
})\right\rangle  \notag \\
&=&\sum_{n=1}^{\infty }\ \frac{2\hbar c}{n}\coth \left[ \frac{\hbar cn\pi }{%
2(b-a)}\right] \sin \!\left[ \frac{n\pi }{b-a}(x-a)\right] \sin \!\left[ 
\frac{n\pi }{b-a}(x^{\prime }-a)\right] \cos \!\left[ \frac{n\pi }{b-a}%
c(t-t^{\prime })\right]  \label{f35}
\end{eqnarray}

The quantum point of view regarding thermal radiation is strikingly
different from the classical viewpoint. The vacuum of the quantum scalar
field is said to involve fluctuations but no quanta, no elementary
excitations, no scalar photons, whereas the thermal radiation field involves
a distinct pattern of scalar photons. \ If the index $m$ is used to label
the normal modes in a one-dimensional box, the quantum expectation values
correspond to an incoherent sum over the expectation values for the fields
for all numbers $n_{m}$ of photons of frequency $\omega _{m}=m\pi c/(b-a)$
with a weighting given by the Boltzmann factor $\exp [-n_{m}\hbar \omega
_{m}/(k_{B}T)].$ Thus the quantum two-point field correlation function for
our example involving a box in one spatial dimension is given by\cite%
{Connection} 
\begin{align}
& \left\langle |(1/2)\{\overline{\phi }(ct,x\mathbf{)}\overline{\phi }%
(ct^{\prime },x^{\prime })+\overline{\phi }(ct^{\prime },x^{\prime })%
\overline{\phi }(ct,x\mathbf{)\}|}\right\rangle _{T}  \notag \\
& =\sum_{m=1}^{\infty }\dsum\limits_{n_{m}=0}^{\infty }\frac{1}{Z\{\hbar
c\pi /[(b-a)k_{B}T]\}}\exp \left[ \frac{-n_{m}\hbar cm\pi }{(b-a)k_{B}T}%
\right]  \notag \\
& \times \left\langle n_{m}|(1/2)\{\overline{\phi }(ct,\mathbf{r)}\overline{%
\phi }(ct^{\prime },\mathbf{r}^{\prime })+\overline{\phi }(ct^{\prime },%
\mathbf{r}^{\prime })\overline{\phi }(ct,\mathbf{r)\}|}n_{m}\right\rangle 
\notag \\
& =\sum_{m=1}^{\infty }\ \frac{2\hbar c}{m}\coth \left[ \frac{\hbar cm\pi }{%
2(b-a)}\right] \sin \!\left[ \frac{m\pi }{b-a}(x-a)\right] \sin \!\left[ 
\frac{m\pi }{b-a}(x^{\prime }-a)\right] \cos \!\left[ \frac{m\pi }{b-a}%
c(t-t^{\prime })\right]  \label{f36}
\end{align}%
where we have noted that 
\begin{equation}
\frac{1}{2}\coth \frac{x}{2}=\frac{\dsum\limits_{n=0}^{\infty }(n+1/2)\exp
[-nx]}{\dsum\limits_{n=0}^{\infty }\exp [-nx]}  \label{f37}
\end{equation}%
and have defined%
\begin{equation}
Z(x)=\dsum\limits_{n=0}^{\infty }\exp [-nx]  \label{f38}
\end{equation}%
Thus for symmetrized products of quantum fields, the quantum expectation
value in Eq. (\ref{f36}) is in exact agreement with the corresponding
classical average value found in Eq. (\ref{f35}). \ Again the agreement
holds for higher order correlation functions provided the quantum operator
order is completely symmetrized.\cite{Connection}

The agreement between the classical and quantum correlation functions
remains in the limits of a large box $b\rightarrow \infty $ analogous to the
transition from Eq. (\ref{e24}) over to Eq. (\ref{f26}) and in the removal
of the left-hand mirror to negative spatial infinity as in the transition
from Eq. (\ref{f26}) over to Eq. (\ref{f27}).

It should be emphasized that although there is complete agreement between
the correlation functions arising in classical theory and the symmetrized
expectation values in quantum theory, the interpretations in terms of
fluctuations arising from classical wave interference or in terms of
fluctuations arising from the presence of photons are completely different
between the theories.\cite{1969} \ The contrast in interpretations and
indeed in predictions becomes even more striking when an accelerating
coordinate frame is involved.

\section{Radiation in a Rindler Frame}

\subsection{Rindler Coordinate Frame}

Although there is close agreement between classical and quantum field
theories in an inertial frame, the two theories part company in noninertial
frames. \ The noninertial frame which we will consider in this article is a
Rindler coordinate frame accelerating through Minkowski spacetime in two
spacetime dimensions.\cite{Rindler}\cite{RindlerF} \ If the coordinates of a
spacetime point in an inertial frame are given by $(ct,x)$, then the
coordinates $(\eta ,\xi )$ of the spacetime point in the Rindler frame which
is at rest with respect to the inertial frame at time $t=0=\eta $ are given
by%
\begin{equation}
ct=\xi \sinh \eta  \label{f39}
\end{equation}%
\begin{equation}
x=\xi \cosh \eta  \label{f40}
\end{equation}%
with $-\infty <\eta <\infty $, and $0<\xi $. Using the relation $\cosh
^{2}\eta -\sinh ^{2}\eta =1$, it follows that a point with fixed spatial
coordinate $\xi $ in the Rindler frame has coordinates $x_{\xi }(t)$ in the
inertial frame given by 
\begin{equation}
x_{\xi }(t)=\xi \cosh \eta =(\xi ^{2}+\xi ^{2}\sinh \eta )^{1/2}=(\xi
^{2}+c^{2}t^{2})^{1/2}  \label{f41}
\end{equation}%
and so moves with acceleration $a_{\xi }=d^{2}x/dt^{2}=c^{2}/\xi $ at time $%
t=0,$ and indeed in the Rindler frame has constant proper acceleration 
\begin{equation}
a_{\xi }=c^{2}/\xi  \label{f42}
\end{equation}%
at all times. Thus for large coordinates $\xi ,$ the acceleration $a_{\xi }$
becomes small whereas for small $\xi $, the proper acceleration diverges.
The point $\xi =0$ is termed the "event horizon" for the Rindler coordinate
frame.

The metric in the Rindler frame can be obtained from Eqs. (\ref{f39}) and (%
\ref{f40}) as%
\begin{equation}
ds^{2}=dt^{2}-dx^{2}=\xi ^{2}d\eta ^{2}-d\xi ^{2}  \label{metric}
\end{equation}%
It is clear from this expression that the time coordinate $\eta $ in the
Rindler frame is not a geodesic coordinate. \ Indeed, the geodesic
separation between two spacetime points which takes the form $%
c^{2}(t-t^{\prime })^{2}-(x-x^{\prime })^{2}$ in the geodesic coordinates of
the inertial frame becomes in Rindler coordinates 
\begin{eqnarray}
c^{2}(t-t^{\prime })^{2}-(x-x^{\prime })^{2} &=&(\xi \sinh \eta -\xi
^{\prime }\sinh \eta ^{\prime })^{2}-(\xi \cosh \eta -\xi ^{\prime }\cosh
\eta ^{\prime })^{2}  \notag \\
&=&2\xi \xi ^{\prime }\cosh (\eta -\eta ^{\prime })-\xi ^{2}-\xi ^{\prime 2}
\label{R1}
\end{eqnarray}

\subsection{Normal Modes in a Box in a Rindler Frame}

We now consider the spectrum of random radiation as observed in the Rindler
frame. First we obtain the radiation normal modes. The wave equation (\ref%
{e3}) in an inertial frame can be transformed to the wave equation in the
Rindler frame by using the transformations (\ref{f39}) and (\ref{f40})
together with the scalar behavior of the field $\phi $ under a coordinate
transformation. The scalar field takes the same value in any coordinate
frame. Thus the field $\varphi (\eta ,\xi )$ in the Rindler frame is equal
to the field $\phi (ct,x)$ in the inertial frame at the same spacetime
point, 
\begin{equation}
\varphi (\eta ,\xi )=\phi (ct,x)=\phi (\xi \sinh \eta ,\xi \cosh \eta ).
\label{e61}
\end{equation}%
If we use the usual rules for partial derivatives, we find that Eq.~(\ref{e3}%
) becomes in the Rindler frame 
\begin{equation}
\left( \frac{\partial ^{2}\varphi }{\partial \xi ^{2}}\right) +\frac{1}{\xi }%
\left( \frac{\partial \varphi }{\partial \xi }\right) -\frac{1}{\xi ^{2}}%
\left( \frac{\partial ^{2}\varphi }{\partial \eta ^{2}}\right) =0.
\label{e62}
\end{equation}%
The solutions of Eq.~(\ref{e62}) take the form $H(\ln \xi \pm \eta )$ where $%
H$ is an arbitrary function. Thus, whereas the general solution of the
scalar wave equation (\ref{e3}) in an inertial frame is $\phi
(ct,x)=h_{+}(x-ct)+h_{-}(x+ct)$ where $h_{+}$ and $h_{-}$ are arbitrary
functions, the general solution in a Rindler frame is $\varphi (\eta ,\xi
)=H_{+}(\ln \xi -\eta )+H_{-}(\ln \xi +\eta )$ where $H_{+}$ and $H_{-}$ are
arbitrary functions. The normal mode solutions of the wave equation in the
Rindler frame for a box extending from $0<\xi =a$ to $\xi =b$ with Dirichlet
boundary conditions can be obtained by separation of variables and expressed
as a time-Fourier series 
\begin{equation}
\varphi _{n}(\eta ,\xi )=\mathcal{F}_{n}\left( \frac{2}{\ln (b/a)}\right)
^{1/2}\sin \left[ \frac{n\pi }{\ln (b/a)}\ln \left( \frac{\xi }{a}\right) %
\right] \cos \left[ \frac{n\pi }{\ln (b/a)}\eta +\theta _{n}\right] ,\quad
(n=1,2,3\ldots ),  \label{e63}
\end{equation}%
where $\mathcal{F}_{n}$ is the amplitude of the normal mode and the spatial
functions 
\begin{equation}
\psi _{n}(\xi )=\left( \frac{2}{\ln (b/a)}\right) ^{1/2}\sin \left[ \frac{%
n\pi }{\ln (b/a)}\ln \left( \frac{\xi }{a}\right) \right] ,  \label{e64}
\end{equation}%
arise from a Sturm-Liouville system\cite{Sturm-L} and form a complete
orthonormal set with weight $1/\xi $ on the interval $a<\xi <b$ . Thus we
find 
\begin{align}
\int_{a}^{b}\frac{d\xi }{\xi }\psi _{n}(\xi )\psi _{m}(\xi )& =\!\int_{a}^{b}%
\frac{d\xi }{\xi }\frac{2}{\ln (b/a)}\sin \left[ \frac{n\pi }{\ln (b/a)}\ln
\left( \frac{\xi }{a}\right) \right] \sin \left[ \frac{m\pi }{\ln (b/a)}\ln
\left( \frac{\xi }{a}\right) \right]  \notag \\
& =\!\int_{v=0}^{v=\pi }\frac{\ln (b/a)}{\pi }dv\frac{2}{\ln (b/a)}\sin
nv\sin mv=\delta _{nm},  \label{e65}
\end{align}%
where we have used the substitution $v=[\pi \ln (\xi /a)]/\ln (b/a)~$in
evaluating the integral. For a radiation normal mode, the Rindler time
parameter $\eta $ agrees with all local clocks when adjusted by $\xi $, and
thus the time $\tau =\xi \eta $ gives the proper time of a clock located at
fixed Rindler spatial coordinate $\xi $.

For time-stationary random radiation in the Rindler frame with an unknown
time-spectral amplitude $\mathcal{F}_{n}$, the field $\varphi (\eta ,\xi )$
can be written as a sum over the normal modes $\varphi _{n}(\eta ,\xi )$ in
Eq.~(\ref{e63}) with random phases $\theta _{n}$ distributed randomly over
the interval $[0,2\pi )$ and distributed independently for each value of $n$%
\begin{equation}
\varphi _{box}(\eta ,\xi )=\dsum_{n=1}^{\infty }\mathcal{F}_{n}\left( \frac{2%
}{\ln (b/a)}\right) ^{1/2}\sin \left[ \frac{n\pi }{\ln (b/a)}\ln \left( 
\frac{\xi }{a}\right) \right] \cos \left[ \frac{n\pi }{\ln (b/a)}\eta
+\theta _{n}\right]  \label{e66}
\end{equation}%
Then the two-field correlation function is obtained in analogy with Eqs.~(%
\ref{f24})--(\ref{e24})%
\begin{eqnarray}
&&\left\langle \varphi _{box}(\eta ,\xi )\varphi _{box}(\eta ^{\prime },\xi
^{\prime })\right\rangle  \notag \\
&=&\dsum_{n=1}^{n=\infty }\mathcal{F}_{n}^{2}\left( \frac{1}{\ln (b/a)}%
\right) \sin \left[ \frac{n\pi }{\ln (b/a)}\ln \left( \frac{\xi }{a}\right) %
\right] \sin \left[ \frac{n\pi }{\ln (b/a)}\ln \left( \frac{\xi ^{\prime }}{a%
}\right) \right] \cos \left[ \frac{n\pi (\eta -\eta ^{\prime })}{\ln (b/a)}%
\right]  \label{e67}
\end{eqnarray}%
For a large box $b\rightarrow \infty $, The normal mode frequencies $\kappa
_{n}=n\pi /\ln (b/a)$ become continuous and the sum in Eq. (\ref{e67})
becomes the integral for the correlation function for a mirror at the
left-hand edge $\xi =a$ of the box%
\begin{equation}
\left\langle \varphi _{mirror}(\eta ,\xi )\varphi _{mirror}(\eta ^{\prime
},\xi ^{\prime })\right\rangle =\frac{1}{\pi }\dint_{\kappa =0}^{\infty
}d\kappa \mathcal{F}^{2}(\kappa )\sin \left[ \kappa \ln \left( \frac{\xi }{a}%
\right) \right] \sin \left[ \kappa \ln \left( \frac{\xi ^{\prime }}{a}%
\right) \right] \cos \left[ \kappa (\eta -\eta ^{\prime })\right]
\label{e68}
\end{equation}%
The expression (\ref{e68}) can be rewritten in the form

\begin{eqnarray}
&&\left\langle \varphi _{mirror}(\eta ,\xi )\varphi _{mirror}(\eta ^{\prime
},\xi ^{\prime })\right\rangle  \notag \\
&=&\frac{1}{2\pi }\dint_{\kappa =0}^{\infty }d\kappa \mathcal{F}^{2}(\kappa
)\cos \left[ \kappa (\ln \xi -\ln \xi ^{\prime })\right] \cos \left[ \kappa
(\eta -\eta ^{\prime })\right]  \notag \\
&&-\frac{1}{2\pi }\dint_{\kappa =0}^{\infty }d\kappa \mathcal{F}^{2}(\kappa
)\cos \left[ \kappa (\ln \xi +\ln \xi ^{\prime }-2\ln a)\right] \cos \left[
\kappa (\eta -\eta ^{\prime })\right]  \label{e69}
\end{eqnarray}%
In the limit $a\rightarrow 0$ in which the mirror at $\xi =a$ is moved to
the event horizon, the last integral in Eq. (\ref{e69}) involves a rapidly
oscillating cosine function; it can be taken to vanish when considering the
time derivative at $\xi =\xi ^{\prime }$. \ Thus we find the free-space
expression 
\begin{equation}
\left\langle \varphi (\eta ,\xi )\partial _{\eta ^{\prime }}\varphi (\eta
^{\prime },\xi ^{\prime })\right\rangle _{\xi =\xi ^{\prime }}=\frac{1}{4\pi 
}\dint_{-\infty }^{\infty }d\kappa \mathcal{F}^{2}(\kappa )\kappa \sin \left[
\kappa (\eta -\eta ^{\prime })\right]  \label{e70}
\end{equation}%
\ where the spectral amplitude $\mathcal{F}(\kappa )$ of the random
radiation is still unspecified.

\subsection{Classical Zero-Point Radiation in the Rindler-Frame Box}

It was noted earlier that the spectrum of classical zero-point radiation
follows from the assumed symmetry properties of the vacuum. \ Thus the
spectrum of random classical radiation in empty space is assumed to be
featureless; the two-point correlation function can depend upon only the
geodesic separation (and its coordinate derivatives) between the spacetime
points. This dependence upon the geodesic separation has been exhibited in
earlier articles for the relativistic scalar and electromagnetic fields in
four spacetime dimensions.\cite{conformal}\cite{Interpret} For the example
of two spacetime dimensions used in the present article, the derivative
correlation functions (\ref{f28}) and (\ref{f29}) involve the partial
derivatives of the logarithm of the spacetime separation $\left\vert
c^{2}(t-t^{\prime })^{2}-(x-x^{\prime })^{2}\right\vert $ between the
spacetime points $(ct,x)$ and $(ct^{\prime },x^{\prime }).$ \ 

In classical theory, the zero-point radiation is physically present. \ There
is no notion of "virtual" photons which may come into and then out of
existence. \ Thus in empty space, the spectrum of radiation which is found
in the Rindler frame follows directly by tensor transformation from the
radiation found in the inertial frame. \ We find for a scalar field that the
correlation function is the same in the inertial frame and the Rindler frame
for the same spacetime points%
\begin{equation}
\left\langle \varphi (\eta ,\xi )\varphi (\eta ^{\prime },\xi ^{\prime
})\right\rangle =\left\langle \phi (ct,x)\phi (ct,x)\right\rangle
=\left\langle \phi (\xi \sinh \eta ,\xi \cosh \eta )\phi (\xi ^{\prime
}\sinh \eta ^{\prime },\xi ^{\prime }\cosh \eta ^{\prime })\right\rangle
\label{k11}
\end{equation}%
However, it is clear from this equation (\ref{k11}) that the functional
dependence of the correlation function upon $\xi ,\xi ^{\prime },\eta ,\eta
^{\prime }$ will in general be quite different from the dependence upon $%
x,x^{\prime },t,t^{\prime }$ since from Eq. (\ref{R1}), the geodesic
separation takes the form $c^{2}(t-t^{\prime })^{2}-(x-x^{\prime })^{2}=2\xi
\xi ^{\prime }\cosh (\eta -\eta ^{\prime })-\xi ^{2}-\xi ^{\prime 2},$ and
the Rindler frame time parameter $\eta $ is not a geodesic coordinate. \ In
empty space, the closed form expressions for the spatial derivatives of the
correlation function in the Rindler frame follow from Eqs. (\ref{f28}), (\ref%
{f29}), (\ref{R1})and (\ref{k11}) as%
\begin{equation}
\left\langle \varphi _{0}(\eta ,\xi )\partial _{\eta ^{\prime }}\varphi
_{0}(\eta ^{\prime },\xi ^{\prime })\right\rangle =\partial _{\eta ^{\prime
}}\left\{ -\hbar c\ln \left\vert 2\xi \xi ^{\prime }\cosh (\eta -\eta
^{\prime })-\xi ^{2}-\xi ^{\prime 2}\right\vert \right\}  \label{k11a}
\end{equation}%
\begin{equation}
\left\langle \varphi _{0}(\eta ,\xi )\partial _{\xi ^{\prime }}\varphi
_{0}(\eta ^{\prime },\xi ^{\prime })\right\rangle =\partial _{\xi ^{\prime
}}\left\{ -\hbar c\ln \left\vert 2\xi \xi ^{\prime }\cosh (\eta -\eta
^{\prime })-\xi ^{2}-\xi ^{\prime 2}\right\vert \right\}  \label{k11b}
\end{equation}

The time-spectrum found in the Rindler frame may be obtained by taking the
singular Fourier sine transform of the time correlation at a single spatial
coordinate $\xi =\xi ^{\prime }.$ \ Thus from Eq. (\ref{e70}) and (\ref{k11a}%
), we find for the spectral function corresponding to classical zero-point
radiation\cite{GR}%
\begin{eqnarray}
\mathcal{F}_{0}^{2}(\kappa ) &=&\frac{4}{\kappa }\dint_{0}^{\infty }d(\eta
-\eta ^{\prime })\sin [\kappa (\eta -\eta ^{\prime })]\left\langle \varphi
_{0}(\eta ,\xi )\partial _{\eta ^{\prime }}\varphi _{0}(\eta ^{\prime },\xi
^{\prime })\right\rangle _{\xi =\xi ^{\prime }}  \notag \\
&=&\frac{4}{\kappa }\dint_{0}^{\infty }d(\eta -\eta ^{\prime })\sin [\kappa
(\eta -\eta ^{\prime })]\frac{\hbar c\sinh (\eta -\eta ^{\prime })}{\cosh
(\eta -\eta ^{\prime })-1}  \notag \\
&=&\frac{4\hbar c}{\kappa }\dint_{0}^{\infty }du\sin (\kappa u)\coth \left( 
\frac{u}{2}\right) =\frac{4\hbar c}{\kappa }\pi \coth \left[ \kappa \pi %
\right]  \label{k14}
\end{eqnarray}%
In a Rindler frame box of finite length, this spectral function (\ref{k14})
is restricted to the allowed normal modes $\kappa _{n}=n\pi /\ln (b/a),$ so
that%
\begin{eqnarray}
\varphi _{0box}(\eta ,\xi ) &=&\dsum_{n=1}^{n=\infty }\left( 4\pi \frac{%
\hbar c\ln (b/a)}{n\pi }\coth \left[ \frac{n\pi ^{2}}{\ln (b/a)}\right]
\right) ^{1/2}\left( \frac{2}{\ln (b/a)}\right) ^{1/2}\sin \left[ \frac{n\pi 
}{\ln (b/a)}\ln \left( \frac{\xi }{a}\right) \right]  \notag \\
&&\times \cos \left[ \frac{n\pi }{\ln (b/a)}\eta +\theta _{n}\right]
\label{k15}
\end{eqnarray}%
and the two-point correlation function in the box is given by 
\begin{eqnarray}
&&\left\langle \varphi _{0box}(\eta ,\xi )\varphi _{0box}(\eta ^{\prime
},\xi ^{\prime })\right\rangle  \notag \\
&=&\dsum_{n=1}^{n=\infty }4\pi \frac{\hbar c\ln (b/a)}{n\pi }\coth \left[ 
\frac{n\pi ^{2}}{\ln (b/a)}\right] \left( \frac{1}{\ln (b/a)}\right) \sin %
\left[ \frac{n\pi }{\ln (b/a)}\ln \left( \frac{\xi }{a}\right) \right] \sin %
\left[ \frac{n\pi }{\ln (b/a)}\ln \left( \frac{\xi ^{\prime }}{a}\right) %
\right]  \notag \\
&&\times \cos \left[ \frac{n\pi (\eta -\eta ^{\prime })}{\ln (b/a)}\right]
\label{k16}
\end{eqnarray}

In the limit as $b\rightarrow \infty ,$ corresponding to the right-hand edge
of the box going to positive spatial infinity, the normal mode frequencies $%
\kappa _{n}=n\pi /\ln (b/a)$ become continuous, and the correlation function
(\ref{k16}) becomes that for a mirror at the left-hand edge $\xi =a$ of the
box$,$\ 
\begin{eqnarray}
&&\left\langle \varphi _{0mirror}(\eta ,\xi )\varphi _{0mirror}(\eta
^{\prime },\xi ^{\prime })\right\rangle  \notag \\
&=&4\hbar c\dint_{\kappa =0}^{\infty }\frac{d\kappa }{\kappa }\coth \left[
\kappa \pi \right] \sin \left[ \kappa \ln \left( \frac{\xi }{a}\right) %
\right] \sin \left[ \kappa \ln \left( \frac{\xi ^{\prime }}{a}\right) \right]
\cos \left[ \kappa (\eta -\eta ^{\prime })\right]  \label{k17}
\end{eqnarray}%
This is a convergent integral which can be evaluated as\cite{GR}%
\begin{eqnarray}
&&\left\langle \varphi _{0mirror}(\eta ,\xi )\varphi _{0mirror}(\eta
^{\prime },\xi ^{\prime })\right\rangle  \notag \\
&=&-\hbar c\ln \left\vert \frac{\sinh [\{\ln (\xi /a)-\ln (\xi ^{\prime
}/a)+(\eta -\eta ^{\prime })\}/2]\sinh [\{\ln (\xi /a)-\ln (\xi ^{\prime
}/a)-(\eta -\eta ^{\prime })\}/2]}{\sinh [\{\ln (\xi /a)+\ln (\xi ^{\prime
}/a)+(\eta -\eta ^{\prime })\}/2]\sinh [\{\ln (\xi /a)+\ln (\xi ^{\prime
}/a)-(\eta -\eta ^{\prime })\}/2]}\right\vert  \notag \\
&=&-\hbar c\ln \left\vert \frac{\sinh [\{\ln (\xi /\xi ^{\prime })+(\eta
-\eta ^{\prime })\}/2]\sinh [\{\ln (\xi /\xi ^{\prime })-(\eta -\eta
^{\prime })\}/2]}{\sinh [\{\ln (\xi \xi ^{\prime }/a^{2})+(\eta -\eta
^{\prime })\}/2]\sinh [\{\ln (\xi \xi ^{\prime }/a^{2})-(\eta -\eta ^{\prime
})\}/2]}\right\vert  \label{kk17}
\end{eqnarray}

In the limit where the mirror at $\xi =a$ is moved to the event horizon, $%
a\rightarrow 0$, the correlation function in Eq. (\ref{kk17}) diverges as $%
\hbar c\ln \left\vert \xi \xi ^{\prime }/(a)^{2}\right\vert =\hbar c\ln
\left\vert \xi \xi ^{\prime }\right\vert -\hbar c\ln \left\vert
(a)^{2}\right\vert ,$ which appears similar to the divergence in Eq. (\ref%
{f26a}), except that in previous case $a\rightarrow -\infty $ whereas here $%
a\rightarrow 0.$ \ Just as was done earlier, the divergence can be
eliminated by taking coordinate derivatives. \ In this limit, the
correlation function (\ref{kk17}) should correspond to that for empty space
since as the mirror goes to the event horizon of the Rindler frame, the
phases of waves change very rapidly with distance, and we expect that the
phases of the incident and reflected waves should become uncoupled. \ In the
limit $a\rightarrow 0,$ the correlation function for the mirror (\ref{kk17})
becomes (with divergence-eliminating coordinate derivatives)%
\begin{eqnarray}
&&\partial _{\mu ^{\prime} }\left\langle \varphi _{0}(\eta ,\xi )\varphi _{0}(\eta
^{\prime },\xi ^{\prime })\right\rangle  \notag \\
&=&\partial _{\mu ^{\prime} }\left\{ -\hbar c\ln \left\vert 4\xi \xi ^{\prime }\sinh
[\{\ln (\xi /\xi ^{\prime })+(\eta -\eta ^{\prime })\}/2]\sinh [\{\ln (\xi
/\xi ^{\prime })-(\eta -\eta ^{\prime })\}/2]\right\vert \right\}
\label{kk17a}
\end{eqnarray}%
This expression indeed agrees with the correlation functions for empty space
given in (\ref{k11a}) and (\ref{k11b}) since%
\begin{eqnarray}
&&-4\xi \xi ^{\prime }\sinh [\{\ln (\xi /\xi ^{\prime })+(\eta -\eta
^{\prime })\}/2]\sinh [\{\ln (\xi /\xi ^{\prime })-(\eta -\eta ^{\prime
})\}/2]  \notag \\
&=&2\xi \xi ^{\prime }\cosh (\eta -\eta ^{\prime })-\xi ^{2}-\xi ^{\prime 2}
\label{kk17b}
\end{eqnarray}%
Thus a box with classical zero-point radiation takes on the empty-space
zero-point correlation function when the box is expanded to cover the entire
Rindler spacetime region (the Rindler wedge). \ The presence of any
reflecting walls on the Rindler box becomes ever less important as the walls
recede to the limits of the Rindler region.

If we consider the zero-point correlation function in free space as a
function of space for a single time $\eta =\eta ^{\prime }$ or as a function
of time for a single coordinate $\xi =\xi ^{\prime }$ in the Rindler frame,
then we find the non-vanishing two-point correlations in free space from
Eqs. (\ref{k11a}) and (\ref{k11b}),%
\begin{eqnarray}
\left\langle \varphi _{0}(\eta ,\xi )\partial _{\eta ^{\prime }}\varphi
_{0}(\eta ^{\prime },\xi ^{\prime })\right\rangle _{\xi =\xi ^{\prime }}
&=&2\hbar c\frac{2\sinh [(\eta -\eta ^{\prime })/2]\cosh [(\eta -\eta
^{\prime })/2]}{4\sinh ^{2}[(\eta -\eta ^{\prime })/2]}  \notag \\
&=&\hbar c\coth \left( \frac{\eta -\eta ^{\prime }}{2}\right)  \label{k18}
\end{eqnarray}%
\begin{equation}
\left\langle \varphi _{0}(\eta ,\xi )\partial _{\xi ^{\prime }}\varphi
_{0}(\eta ^{\prime },\xi ^{\prime })\right\rangle _{\eta =\eta ^{\prime }}=%
\frac{2\hbar c}{\xi -\xi ^{\prime }}  \label{k19}
\end{equation}%
and%
\begin{equation}
\left\langle \varphi _{0}(\eta ,\xi )\partial _{\xi ^{\prime }}\varphi
_{0}(\eta ^{\prime },\xi ^{\prime })\right\rangle _{\xi =\xi ^{\prime }}=-%
\frac{\hbar c}{\xi }  \label{k20}
\end{equation}

\subsection{Canonical Quantization in a Rindler-Frame Box}

Quantum theory regards canonical quantization as a fundamental procedure
which can be followed in any box, no matter whether the box is at rest in an
inertial frame or is at rest in a noninertial coordinate frame. \ Thus for a
box in a Rindler frame, the quantum field can be expressed in a form
parallel to Eq. (\ref{f22}) as%
\begin{eqnarray}
\overline{\varphi }_{box}(\eta ,\xi ) &=&\dsum_{n=1}^{n=\infty }\left( 8\pi
\hbar c\frac{\ln (b/a)}{n\pi }\right) ^{1/2}\left( \frac{2}{\ln (b/a)}%
\right) ^{1/2}\sin \left[ \frac{n\pi }{\ln (b/a)}\ln \left( \frac{\xi }{a}%
\right) \right]  \notag \\
&&\times \frac{1}{2}\left\{ \overline{b}_{n}\exp \left[ i\frac{n\pi }{\ln
(b/a)}\eta \right] +\overline{b}_{n}^{+}\exp \left[ -i\frac{n\pi }{\ln (b/a)}%
\eta \right] \right\}  \label{k1}
\end{eqnarray}%
where $\overline{b}_{n}$ and $\overline{b}_{n}^{+}$ are the annihilation and
creation operators for particles in the Rindler-frame box. \ Notice that the
amplitude appearing in the sum is the same factor involving the square root
of $8\pi \hbar c$ times the wave number, \ just as in Eq. (\ref{f22}) in the
inertial frame in empty space. \ In contrast, the classical theory involves
the amplitude factor $\mathcal{F}_{0}(\kappa )$ given in Eq. (\ref{k14}) in
order to compensate for the fact that the time coordinate for the normal
modes is not a geodesic coordinate. In quantum theory, there is a
Rindler-frame vacuum state $|0_{R}>$ which is annihilated by the Rindler
operator $\overline{b}_{n}.$ \ The two-point Rindler-vacuum expectation
value for the symmetrized product of the field operators gives the result
parallel to Eq. (\ref{f23}) as%
\begin{eqnarray}
&<&0_{R}|\frac{1}{2}\{\overline{\varphi }_{box}(\eta ,\xi )\overline{\varphi 
}_{box}(\eta ^{\prime },\xi ^{\prime })+\overline{\varphi }_{box}(\eta
^{\prime },\xi ^{\prime })\overline{\varphi }_{box}(\eta ,\xi )\}|0_{R}> 
\notag \\
&=&\dsum_{n=1}^{n=\infty }\frac{4\hbar c}{n}\sin \left[ \frac{n\pi }{\ln
(b/a)}\ln \left( \frac{\xi }{a}\right) \right] \sin \left[ \frac{n\pi }{\ln
(b/a)}\ln \left( \frac{\xi }{a}\right) \right] \cos \left[ \frac{n\pi (\eta
-\eta ^{\prime })}{\ln (b/a)}\right]  \label{k2}
\end{eqnarray}%
In the limit as $b\rightarrow \infty ,$ this expression becomes the
Rindler-vacuum expectation value for the situation of continuous normal mode
frequencies $\kappa _{n}=n\pi /\ln (b/a)$ and a mirror at $\xi =a,$
analogous to Eqs. (\ref{f26}), and (\ref{f26a}), 
\begin{eqnarray}
&<&0_{R}|\frac{1}{2}\{\overline{\varphi }_{mirror}(\eta ,\xi )\overline{%
\varphi }_{mirror}(\eta ^{\prime },\xi ^{\prime })+\overline{\varphi }%
_{mirror}(\eta ^{\prime },\xi ^{\prime })\overline{\varphi }_{mirror}(\eta
,\xi )\}|0_{R}>  \notag \\
&=&2\hbar c\dint_{0}^{\infty }\frac{d\kappa }{\kappa }\sin \left[ \kappa \ln
\left( \frac{\xi }{a}\right) \right] \sin \left[ \kappa \ln \left( \frac{\xi
^{\prime }}{a}\right) \right] \cos [\kappa (\eta -\eta ^{\prime })]  \notag
\\
&=&-\hbar c\ln \left\vert \frac{[\{\ln (\xi /a)-\ln (\xi ^{\prime
}/a)\}-c(t-t^{\prime })][\{\ln (\xi /a)-\ln (\xi ^{\prime
}/a)\}-c(t-t^{\prime })]}{[\{\ln (\xi /a)+\ln (\xi ^{\prime
}/a)\}-c(t-t^{\prime })][\{\ln (\xi /a)+\ln (\xi ^{\prime
}/a)\}-c(t-t^{\prime })]}\right\vert  \notag \\
&=&-\hbar c\ln \left\vert \frac{\{\ln (\xi /\xi ^{\prime
})\}^{2}-c^{2}(t-t^{\prime })^{2}}{\{\ln (\xi /a)+\ln (\xi ^{\prime
}/a)\}^{2}-c^{2}(t-t^{\prime })^{2}}\right\vert  \label{k3}
\end{eqnarray}%
In the limit $a\rightarrow 0$ that the mirror is moved to the event horizon,
the expectation value for the quantum fields in the Rindler vacuum becomes
divergent as $2\hbar c\ln [2\ln (a)].$ \ Again the divergence can be
eliminated by taking coordinate derivatives%
\begin{eqnarray}
\partial _{\mu } &<&0_{R}|\frac{1}{2}\{\overline{\varphi }(\eta ,\xi )%
\overline{\varphi }(\eta ^{\prime },\xi ^{\prime })+\overline{\varphi }(\eta
^{\prime },\xi ^{\prime })\overline{\varphi }(\eta ,\xi )\}|0_{R}>  \notag \\
&=&\partial _{\mu }\left\{ -\hbar c\ln \left\vert \{\ln (\xi /\xi ^{\prime
})\}^{2}-c^{2}(t-t^{\prime })^{2}\right\vert \right\}  \label{kk4}
\end{eqnarray}

\ Thus we obtain%
\begin{eqnarray}
&<&0_{R}|\frac{1}{2}\{\overline{\varphi }(\eta ,\xi )\partial _{\eta
^{\prime }}\overline{\varphi }(\eta ^{\prime },\xi ^{\prime })+\partial
_{\eta ^{\prime }}\overline{\varphi }(\eta ^{\prime },\xi ^{\prime })%
\overline{\varphi }(\eta ,\xi )\}|0_{R}>  \notag \\
&=&2\hbar c\frac{(\eta -\eta ^{\prime })}{(\eta -\eta ^{\prime })^{2}-(\ln
\xi -\ln \xi ^{\prime })^{2}}  \label{k6}
\end{eqnarray}%
and%
\begin{eqnarray}
&<&0_{R}|\frac{1}{2}\{\overline{\varphi }(\eta ,\xi )\partial _{\xi ^{\prime
}}\overline{\varphi }(\eta ^{\prime },\xi ^{\prime })+\partial _{\xi
^{\prime }}\overline{\varphi }(\eta ^{\prime },\xi ^{\prime })\overline{%
\varphi }(\eta ,\xi )\}|0_{R}>  \notag \\
&=&2\hbar c\frac{(\ln \xi -\ln \xi ^{\prime })}{(\eta -\eta ^{\prime
})^{2}-(\ln \xi -\ln \xi ^{\prime })^{2}}  \label{k7}
\end{eqnarray}%
If we consider the spatial dependence at a single time and the time
dependence at a single spatial point, we find for the symmetrized
expectation value for the Rindler vacuum that the non-vanishing values from
Eqs. (\ref{k6}) and (\ref{k7}) are 
\begin{equation}
<0_{R}|\frac{1}{2}\{\overline{\varphi }(\eta ,\xi )\partial _{\eta ^{\prime
}}\overline{\varphi }(\eta ^{\prime },\xi ^{\prime })+\partial _{\eta
^{\prime }}\overline{\varphi }(\eta ^{\prime },\xi ^{\prime })\overline{%
\varphi }(\eta ,\xi )\}|0_{R}>_{\xi =\xi ^{\prime }}=2\hbar c\frac{1}{(\eta
-\eta ^{\prime })}  \label{k8}
\end{equation}%
and%
\begin{equation}
<0_{R}|\frac{1}{2}\{\overline{\varphi }(\eta ,\xi )\partial _{\xi ^{\prime }}%
\overline{\varphi }(\eta ^{\prime },\xi ^{\prime })+\partial _{\xi ^{\prime
}}\overline{\varphi }(\eta ^{\prime },\xi ^{\prime })\overline{\varphi }%
(\eta ,\xi )\}|0_{R}>_{\eta =\eta ^{\prime }}=2\hbar c\frac{1}{(\ln \xi -\ln
\xi ^{\prime })}  \label{k9}
\end{equation}%
The Rindler vacuum expectation value in (\ref{k8}) with its dependence upon
the inverse time separation is analogous to the free-space inertial frame
vacuum expectation value (\ref{f32}) in an inertial frame. However, the
Rindler vacuum expectation value (\ref{k9}) with its logarithmic dependence
on $\xi $ and $\xi ^{\prime }$ has no analogue in an inertial frame. \ The
"Rindler vacuum" is different from the "Minkowski vacuum" under canonical
quantization.

\subsection{Contrasting Classical-Quantum Viewpoints in a Rindler Frame}

Although the classical zero-point correlation functions and the quantum
symmetrized vacuum expectation values agree in inertial frames, they are no
longer in agreement in non-inertial frames. \ The vacuum states arise from
very different concepts in the classical and the quantum theories. \ The
essential feature of classical zero-point radiation is that the spectrum of
random radiation is featureless. \ Therefore in empty space, classical
zero-point radiation depends only upon the geodesic separation of the field
points. \ The spectrum obtained from the continuous frequencies of empty
space is then restricted to the allowed normal mode frequencies in a box of
finite size. In the limit where the sides of the box are moved to the limits
of the spacetime, the spectrum in the box becomes that of empty space. \
Thus a box with walls at rest in an inertial frame and a box at rest with
respect to the coordinates of a Rindler frame have very different normal
modes, and the spectral amplitudes are readjusted to reflect the change from
a geodesic to non-geodesic time coordinate. \ In terms of a geodesic time
coordinate such as appears in an inertial frame, the spectrum of zero-point
radiation is given by $f_{0}^{2}(k)=4\pi \hbar c/|k|$ where the constant is
chosen to give an energy ($1/2)\hbar c|k|$ per normal mode. \ In terms of
the non-geodesic time coordinate $\eta $ appearing in a Rindler frame, the
spectrum of zero-point radiation is given by $\mathcal{F}_{0}^{2}(\kappa
)=(4\pi \hbar c/\kappa )\coth (\pi \kappa )$ . \ If the walls of the box are
moved to the limits of the Rindler wedge, the random radiation in the
Rindler space is exactly that of the inertial space. \ The classical vacuum
is unique.

In complete contrast, the vacuum of quantum field theory arises from a
prescriptive process which takes no account of the spacetime metric. \ In
any box, the amplitude for the normal modes is fixed, and annihilation and
creation operators are introduced for the positive and negative time
aspects. \ Thus a box with walls at rest in an inertial frame and a box at
rest with respect to the coordinates of a Rindler frame have very different
normal modes but the same spectral amplitude, and accordingly have very
different vacuum states. \ If the walls of the Rindler box are moved out to
the limits of the Rindler spacetime wedge, the quantum fluctuations
associated with the Rindler vacuum state remain quite different from the
quantum fluctuations associated with the inertial vacuum state. \ The
"Rindler vacuum" is different from the "Minkowski vacuum" even for a large
box. \ There is a non-uniqueness for the quantum vacuum in non-inertial
frames.

Of course, one can apply tensor transformations to the vacuum expectation
values of the symmetrized quantum operators which were found in an inertial
frame. \ Since the symmetrized quantum expectation values agree exactly with
the corresponding classical correlation functions in an inertial frame, we
obtain exactly the same expressions (\ref{k16})-(\ref{k20}) as found for the
classical correlation functions in the Rindler frame. \ The spatial
dependence on the geodesic coordinate $\xi $ found in Eq. (\ref{k19}) for
the correlation function at a single time $\eta =\eta ^{\prime }$ agrees
exactly with that found in the corresponding expression (\ref{f33}) in an
inertial frame (for $x=\xi ,$ $x^{\prime }=\xi ^{\prime })$, as we indeed
expect since a fixed time $\eta =\eta ^{\prime }$ corresponds to a single
time $t=t^{\prime }$ in the momentarily comoving inertial reference frame,
and all inertial frames have the same correlation functions for zero-point
radiation. \ The absence of any spatial correlation length in Eq. (\ref{k19}%
) corresponds to zero-temperature $T=0.$\ However, the time dependence in
Eq. (\ref{k18}) for the correlation function at a single spatial coordinate $%
\xi =\xi ^{\prime }$ is quite different from the time dependence (\ref{f32})
found in an inertial frame. \ Indeed, The appearance of the hyperbolic
cotangent function for the time-Fourier spectrum in Eq. (\ref{k14}) has led
some physicists to speak of the "thermal effects of acceleration through the
vacuum"\cite{Fulling}\cite{Davies}\cite{Unruh}\cite{Milonni}\cite{Crispino}
with temperature $T=\hbar a/(2\pi ck_{B}).$ \ After all, the hyperbolic
cotangent function appeared in Eq. (\ref{f34}) for the spectrum of thermal
radiation in an inertial frame. \ Thus the spectra in the Rindler frame can
be used to suggest either finite temperature $T=\hbar a/(2\pi ck_{B})$ or
zero-temperature $T=0$ depending upon one's point of view. \ This ambiguity
arises precisely because the Rindler frame is not an inertial frame and the
Rindler time parameter $\eta $ is not a geodesic coordinate. Indeed one may
inquire as to just what spectrum corresponds to thermal radiation in a
noninertial frame. \ Within classical physics, this question has been
discussed in connection with time-dilating conformal transformations which
allow us to derive the Planck spectrum from the structure of relativistic
spacetime.\cite{twoD}\cite{Derivation}\cite{structure}

Despite the classical-quantum agreement of the tensor-transformed inertial
expectation values, the quantum viewpoint is more complicated since quantum
theory introduces a new vacuum state associated with canonical quantization
in the Rindler frame. \ Canonical quantization within a box in a Rindler
frame leads to field fluctuations which are quite different from those found
from quantization in an inertial frame. \ Thus the time dependence of the
symmetrized Rindler vacuum expectation value at a single spatial coordinate
in (\ref{k8}) (with its inverse time dependence) is indeed analogous to the
inverse time dependence found in (\ref{f32}) for the inertial frame. \
However, the logarithmic spatial dependence of the Rindler vacuum
expectation value at a single time in (\ref{k9}) is quite different from
that given in (\ref{f33}) for the inertial frame. \ Thus the quantum vacuum
in a Rindler frame has quite different properties from the quantum vacuum in
an inertial frame. Indeed, over 30 years ago, Fulling called attention to
this "Nonuniqueness of Canonical Quantization in Riemannian Space-Time."\cite%
{Fulling} \ 

And what is the physical meaning of the "Rindler vacuum state" which is
different from the familiar "Minkowski vacuum state"? \ According to some
quantum field theorists,\cite{Ref1} the vacuum is established by the walls
of the box which confine the radiation. \ If the walls of the box are
established at temperature $T=0$ in the inertial frame vacuum and then the
box has its acceleration slowly increased to the final acceleration, its
interior will be in the Rindler vacuum. \ On the other hand, if the box at
temperature $T=0$ in the inertial frame is suddenly accelerated, the box
will contain Rindler excitations corresponding to the Fulling-Davies-Unruh
temperature $T=\hbar a/(2\pi ck_{B})$ as measured in the Rindler frame where
the Rindler vacuum is the lowest energy state.\cite{Ref1}

The classical theory with zero-point radiation lends no support to this
quantum interpretation. \ The classical vacuum state involving classical
zero-point radiation is unique; its description between any two coordinate
frames is found by tensor transformations. In particular, classical physics
has nothing like the scenario described above for a box of zero-point
radiation which is moved from an inertial to a Rindler frame. According to
classical theory, (except for small Casimir effects) it matters not how the
box of (featureless) classical zero-point radiation is moved from the
inertial frame into the accelerating Rindler frame; the box of radiation
will always correspond to zero-point radiation as described by tensor
transformation from an inertial frame. This statement seems to come as a
surprise to many physicists who are misled by their experience with spectra
involving finite total energy. The invariant result for a box of zero-point
radiation follows from the very special character of the zero-point spectrum
which has no structure other than that which is given to it by the
coordinates associated with the metric of the spacetime.

In an inertial frame in empty space, the zero-point radiation spectrum is
Lorentz invariant and scale invariant; it depends only upon the separation
(and coordinate derivatives) between the two spacetime points measured along
a geodesic between the points.\cite{conformal}\cite{Interpret} \ Perhaps the
reader can obtain a better sense of the special character of zero-point
radiation from the following considerations. We saw in Eqs. (\ref{f28}) and (%
\ref{f29}) that the spectrum of random classical zero-point radiation for
the scalar field in an inertial frame depends upon the logarithm of the
invariant separation between the two spacetime points. Since we are dealing
with a scalar field, the correlation function takes the same value in the
Rindler frame. If we transform the Minkowski coordinates to Rindler
coordinates, as given in Eqs. (\ref{k11a}) and (\ref{k11b}), we find that
the correlation function is time stationary; it depends upon only the time
difference ($\eta -\eta ^{\prime })$ and not on any initial time. There is
no spectrum of \textit{finite} energy density which has such behavior;
time-translation invariance both in all inertial frames and in all Rindler
frames is a property unique to the zero-point radiation spectrum. \ 

The solutions for the wave equations (\ref{e3}) and (\ref{e62}) are unique
for boundary conditions which specify both the function and its first time
derivative a\bigskip t a single time coordinate. \ We can imagine a box of
zero-point radiation which is at rest in an inertial frame and then is
suddenly accelerated so as to remain at the fixed coordinates of a Rindler
frame. \ If we have a box at rest with respect to the coordinates of a
Rindler frame, it will be instantaneously at rest with respect to some
inertial frame. \ Within the classical theory, the zero-point radiation
within the box differs from the zero-point radiation in the inertial frame
by simply the fact that the box modes are restricted to the normal modes of
the box rather than being the continuous modes of empty space. \ As was
proved in our analysis above, the zero-point radiation in a Rindler box
whose walls are moved to the limits of the Rindler wedge is in complete
agreement with the radiation in the empty-space Rindler frame and the
radiation in the empty-space inertial frame. \ Thus the only difference
between the radiation inside the box and the radiation of the empty-space
inertial frame outside the box are the Casimir aspects associated with the
discreteness of the normal mode spectrum of a finite box. \ For a large box,
the zero-point radiation can be accelerated without changing its spectrum.

In work published earlier,\cite{twoD}\cite{Derivation} it has been pointed
out that the Planck spectrum for classical thermal radiation arises
naturally by considering the time-dilation symmetry of thermal radiation in
a Rindler frame. Thus in an inertial frame, a time-dilating conformal
transformation carries thermal radiation at temperature $T$ into thermal
radiation at temperature $\sigma T$ where $\sigma $ is a positive real
number. Under such a transformation, zero-point radiation in an inertial
frame remains zero-point radiation. However, in a Rindler frame, a
time-dilating conformal transformation carries zero-point radiation into
thermal radiation at a non-zero- temperature.\cite{structure} The
perspective from classical physics suggests that the canonical quantization
procedure in a non-inertial frame may be predicting results which have no
realization in nature.

\subsection{Detectors Accelerating through Classical Zero-Point Radiation}

Although during the 1970's there were discussions as to whether or not
acceleration through the quantum Minkowski vacuum turned virtual photons
into real photons, today quantum field theory claims merely that "detectors"
accelerating through the quantum vacuum behave as through they were in a
thermal bath.\cite{Crispino} \ Indeed one quantum theorist has asserted that
on acceleration through the vacuum, "Steaks will cook, eggs will fry."\cite%
{Referee2} \ Of course, there is no experimental basis for such an
assertion. \ And our suggestion is that such an assertion may be wrong. \ 

Quantum theorist often speak of using a very small system which would not be
affected by gravity in order to examine the thermal bath behavior of
mechanical systems.\cite{Ref2} \ Indeed, within classical physics, there are
calculations for \textit{point} harmonic oscillators\cite{Bosc}\cite{Cosc}
and \textit{point} magnetic dipole rotators\cite{Bdip} accelerated through
classical zero-point radiation; these systems indeed take on values for the
average energy as through they were located in an inertial frame in a
thermal bath with temperature $T=\hbar a/(2\pi ck_{B}).$ \ Point systems
respond simply to the time correlation function and so do not sample
anything regarding spatial extent. \ Indeed, by using time-dilating
conformal transformations it can be shown that if we consider only the
correlations in time at a fixed spatial coordinate without measuring
anything involving spatial extent, then we can not separate out the effects
of acceleration from those of non-zero temperature.\cite{Equiv} \ 

However, are \textit{point} mechanical systems reliable indicators of
thermal behavior? \ We suggest that point systems with internal structure
are not relativistic systems and can not be expected to illustrate
accurately the ideas of a relativistic field theory. \ \ \textit{Point}
systems do not exist as relativistic systems except for point masses. \
Point systems (such as a harmonic oscillator of vanishing spatial extent)
which contain potential energy have no mechanism to show the dependence of
the supporting force on the internal potential energy of the system when the
system is located in a gravitational field or in an accelerating coordinate
frame. \ This situation is in complete contrast with electromagnetic systems
of charged particles; such systems must have finite spatial extent and will
be affected by gravity. When the mechanical system contains electromagnetic
energy, then the mechanism for the connecting the supporting force to the
system potential energy in a gravitational field (or in an accelerating
coordinate frame) involves the droop of the electromagnetic field lines.\cite%
{droop} \ However, a mechanical system with electromagnetic potential
energy, such as a classical hydrogen atom, must have finite spatial extent,
and therefore responds to both the temporal and spatial correlation aspects
of the fluctuating field.

Indeed, this question of finite spatial extent has direct relevance to the
arguments\ given previously regarding "sudden" versus "adiabatic"
acceleration of boxes of radiation. \ We can imagine a mechanical system
located at a fixed position in the interior of a box of radiation which is
moved from an inertial frame over to a Rindler frame. \ Within classical
theory, this mechanical system takes on the same value whether alone and
accelerated through the zero-point radiation of a Minkowski frame or whether
at rest inside a (large) box in an accelerating Rindler frame because the
spectrum of classical zero-point radiation is the same inside or outside the
(large) box. \ However, quantum theory might suggest different behavior for
the mechanical system in these two cases; in the first case the system is
responding to the tensor transformations of the fluctuations of the
Minkowski vacuum and in the second case the system is (presumably)
responding to the fluctuations of the Rindler vacuum. \ Indeed a \textit{%
point} system will simply respond to the local time-fluctuations of the
radiation inside the box. \ This is not true for a hydrogen atom or any
spatially extended relativistic system. \ The field lines of a Coulomb
potential "droop" in a gravitational field and the extent of the "droop" is
a measure of the strength of the gravitational (or acceleration) field. \
The final droop of the field lines of a Coulomb potential in a Rindler frame
has nothing to do with the way in which the potential may have been moved
from an inertial to the Rindler frame. When a \textit{point} harmonic
oscillator is moved up and down in thermal radiation in a gravitational (or
acceleration) field, it can be used to violated fundamental laws of
thermodynamics precisely because it does not readjust to the gravitational
field. \ A hydrogen atom, which is truly a relativistic system, will
readjust to the gravitational field by the droop of the field lines as it is
moved up or down in a Rindler frame. \ Only relativistic systems should be
considered seriously when dealing with relativistic situations. \ It seems
possible that all the claims that acceleration through the vacuum provides a
thermal bath may be in error. \ 

\section{Closing Summary}

Although quantum field theory and classical field theory with classical
zero-point radiation have related vacuum states in inertial frames, the
theories part company in non-inertial frames. \ The vacuum correlation
functions of the classical theory depend upon geodesic separations in the
spacetime whereas the expectation values of the quantum theory depend upon a
canonical quantization procedure which makes no distinction between geodesic
and non-geodesic coordinates. \ The classical vacuum is unique. \ The
non-uniqueness of the quantum vacuum was noted by Fulling over thirty years
ago. \ This contrast invites deeper exploration.\cite{Ref3}


\begin{thebibliography}{99}
\bibitem{Ref1} An anonymous referee for Ref. 6 wrote: "It is also important
to note that, in quantum field theory, if a box with totally reflecting
walls starts off at rest with no acceleration, and with its interior in the
vacuum state, then has its acceleration slowly increased to the final
acceleration, its interior will be in the 'Rindler vacuum', not the thermal
bath. \ Those boundaries to the box make a huge difference." \ A different
referee wrote: "So suppose that we now take the mirror to be inertial for $%
t<0$, and then to accelerate uniformly for $t>0$. Then surely the mirror
again introduces extra correlations and breaks some of the supposed
symmetries of the zero-point radiation. \ If two such mirrors form a box of
size $L$, then after a proper time $L/c$ has elapsed along either mirror,
one would expect even the radiation in the deep interior of the box to be
affected by the mirrors motion."
\bibitem{Ref2} A typical comment is that of an anonymous referee for Ref.
6, "...(again assuming that the body is small enough so that the
gravitational field does not affect it), ..."
\bibitem{Referee2} An anonymous referee for Ref. 6 wrote: "It is important
... to realize that the claim that an accelerated observer see a thermal
bath is based on the fact that such an observer, carrying a thermometer
(which is insensitive in its operation to the presence of a strong
gravitational field) will find it reading a temperature. \ Steaks will cook,
eggs will fry." \ Later this same referee wrote: "I cannot recommend a paper
that denies a quantum field theory direct consequence, namely, that under
uniform acceleration a small thermometer will register a temperature. \
Accelerated enough steaks in Minkowski vacuum will cook, and eggs will fry."
\bibitem{Ref3} This point of view was not shared by a referee for Ref. 6 who
declared, "If there is any disagreement with the standard quantum treatment
then this approach is surely wrong."
\end{thebibliography}
\end{document}